\documentclass[twocolumn,superscriptaddress,aps]{revtex4}
\usepackage{amsmath}
\usepackage{amssymb}
\usepackage{graphicx}
\usepackage{color}

\newcommand{\be}{\begin{equation}}
\newcommand{\ee}{\end{equation}}
\newcommand{\bea}{\begin{eqnarray}}
\newcommand{\eea}{\end{eqnarray}}
\newcommand{\bse}{\begin{subequations}}
\newcommand{\ese}{\end{subequations}}
\newcommand{\sca}{${\rm SrCr_2As_2}$}
\newcommand{\bca}{${\rm BaCr_2As_2}$}
\newcommand{\bma}{${\rm BaMn_2As_2}$}
\newcommand{\bfa}{${\rm BaFe_2As_2}$}
\newcommand{\cfa}{${\rm CaFe_2As_2}$}
\newcommand{\sma}{${\rm SrMn_2As_2}$}
\newcommand{\sfa}{${\rm SrFe_2As_2}$}
\newcommand{\scoa}{${\rm SrCo_2As_2}$}
\newcommand{\ccoa}{${\rm CaCo_{1.86}As_2}$}
\newcommand{\tcs}{${\rm ThCr_2Si_2}$}

\newcommand{\Tn}{$T_{\textrm N}$}

\begin{document}

\title{Itinerant G-type antiferromagnetic order in ${\rm\bf SrCr_2As_2}$}

\author{Pinaki Das}
\email{pdas@ameslab.gov}
\affiliation{Ames Laboratory and Department of Physics and Astronomy, Iowa State University, Ames, Iowa 50011, USA}

\author{N. S. Sangeetha}
\affiliation{Ames Laboratory and Department of Physics and Astronomy, Iowa State University, Ames, Iowa 50011, USA}

\author{George R. Lindemann}
\affiliation{Ames Laboratory and Department of Physics and Astronomy, Iowa State University, Ames, Iowa 50011, USA}

\author{T. W. Heitmann}
\affiliation{The Missouri Research Reactor, University of Missouri, Columbia, Missouri 65211, USA}

\author{A.~Kreyssig}
\author{A. I. Goldman}
\author{R. J. McQueeney}
\author{D. C. Johnston}
\author{D. Vaknin}
\affiliation{Ames Laboratory and Department of Physics and Astronomy, Iowa State University, Ames, Iowa 50011, USA}

\date{\today}

\begin{abstract}

Neutron diffraction and magnetic susceptibility studies of a polycrystalline \sca\ sample reveal that this compound is an itinerant  G-type antiferromagnet below the N${\rm \acute{e}}$el temperature {\Tn} = 590(5)~K with the Cr magnetic moments aligned along the tetragonal $c$ axis. The system remains tetragonal to the lowest measured temperature ($\sim$12 K). The lattice parameter ratio $c/a$ and the magnetic moment saturate at about the same temperature below $\sim$ 200~K, indicating a possible magnetoelastic coupling. The ordered moment, $\mu=1.9(1)~\mu_{\rm B}$/Cr, measured at $T = 12$~K, is significantly reduced compared to its localized value ($4~\mu_{\rm B}$/Cr) due to the itinerant character brought about by the hybridization between the Cr $3d$ and As $4p$ orbitals.

\end{abstract}

\maketitle

\section{Introduction}

Extensive research has been devoted in recent years to iron-based pnictides and chalcogenides due to their intriguing correlated lattice, electronic, magnetic and superconducting properties \cite{Johnston2010, Stewart2011, Scalapino2012, Dagotto2013, Fernandes2014, Hosono2015, Dai2015, Inosov2016, Si2016}.  In particular, comprehensive studies have been conducted on the doped and undoped body-centered tetragonal parent compounds $A{\rm Fe_2As_2}$ ($A$ = Ca, Sr, Ba, Eu) with the \tcs-type structure (122-type compounds). This in turn prompted the search for novel physical properties in other transition-metal based 122-type compounds, such as with Mn/Cr in place of Fe \cite{An2009, Singh2009, Singh2009b, Johnston2011, Antel2012, Calder2014, Zhang2016}, and moreover to ${\rm CaMn_2As_2}$ and ${\rm SrMn_2As_2}$ with the layered trigonal ${\rm CaAl_2Si_2}$-type structure \cite{Sangeetha2016, Das2017}. Experimental and theoretical work on \bca\ with the \tcs-type structure \cite{Pfisterer1980, Pfisterer1983} revealed metallic character, and an itinerant spin-density-wave ground state  \cite{Singh2009a}. The theory also indicated stronger Cr--As covalency than occurs in the Fe--As compounds. \bca\  undergoes G-type antiferromagnetic (AFM) ordering below a transition temperature $T_{\rm N} = 580(10)$~K with moments aligned along the $c$ axis \cite{Filsinger2017}. ARPES measurements indicate reduction in electron correlation effects involving the nominally 3$d^4$ Cr$^{+2}$ cations where the band renormalization is smaller than in ${\rm BaFe_2As_2}$ \cite{Nayak2017, Richard2017}. Additionally, recent electrical resistivity and x-ray diffraction measurements on single and polycrystals of \bca\ under high pressure revealed a tetragonal to collapsed tetragonal (cT) transition at $\sim$~18.5 GPa \cite{Naumov17}. The cT phase has also been manifested in \ccoa\ at ambient pressure \cite{Quirinale13} and in \cfa\ and \scoa\ under high pressures \cite{Goldman09,Jayasekara15}. Measurements on isostructural ${\rm EuCr_2As_2}$ containing divalent Eu cations with spin $S=7/2$ showed this compound to be metallic, with the Cr and Eu sublattices each exhibiting G-type AFM ordering at $T_{\rm N} = 680(40)$~K and 21.0(1)~K, respectively, with the ordered moments on both sublattices aligned along the tetragonal $c$~axis \cite{Paramanik2014, Nandi2016}. The recent discovery of superconductivity in $M_2$Cr$_3$As$_3$ ($M$ = K, Cs, Rb) under ambient pressure \cite{Bao2015,Tang2015a,Tang2015b} and in CrAs under high pressure \cite{Wu2014,Kotegawa2014} sparked more interest in the search for new Cr-As based compounds.

\begin{figure}
\includegraphics{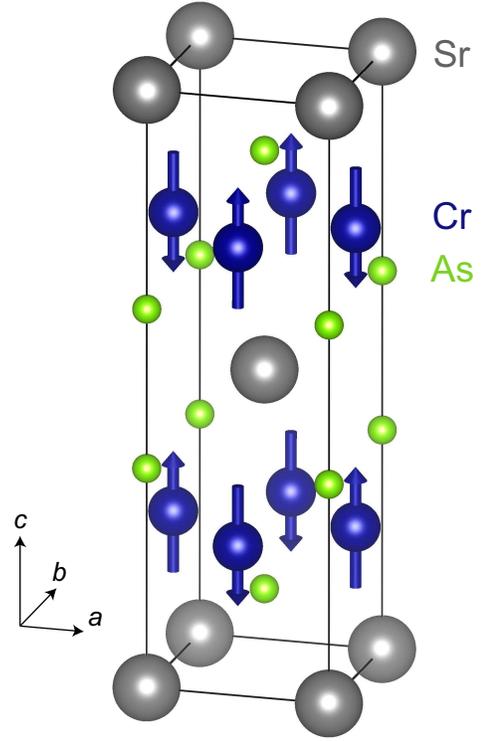}
\caption{(Color online) Chemical and magnetic structures of \sca\ in the magnetically-ordered state. The Cr$^{2+}$ ordered moments are aligned in a G-type arrangement shown by arrows, with antiferromagnetic alignments between all nearest neighbors.}
\label{Fig:MagStructure}
\end{figure}

\sca\ is isostructural to \bca\ \cite{Pfisterer1980, Pfisterer1983}, for which a hint of a magnetic transition at $\sim 165$~K was reported in an early magnetic susceptibility versus temperature study, $\chi(T)$, and attributed to an AFM transition \cite{Pfisterer1983}. This compound is found to be a good metal from $ab$-plane electrical resistivity versus temperature measurements \cite{Sangeetha2017}. Herein, we report neutron diffraction and magnetization studies of a high quality polycrystalline \sca\ sample and show that this compound orders in a G-type AFM structure below {\Tn} = 590(5)~K with the ordered Cr magnetic moments aligned along the tetragonal $c$ axis (see Fig.\ \ref{Fig:MagStructure}). We find no structural distortion down to 12~K but the close resemblance of the temperature variation of the magnetic moment and the lattice parameter ratio, $c/a$, indicates a possible magnetoelastic coupling. The ordered magnetic moment, $\mu=1.9(1)~\mu_{\rm B}$/Cr, is significantly reduced compared to its localized-moment value ($4~\mu_{\rm B}$/Cr$^{2+}$) due to the itinerant character brought about by the spin-dependent hybridization \cite{Singh2009a} between the Cr $3d$ and the As $4p$ orbitals.  This suggests that Cr as a dopant is a stronger scatterer compared to Co or Ni dopants and may explain why superconductivity has not been observed in Cr-doped \bfa\ \cite{Singh2009a,Filsinger2017}.

\section{Experimental Details}

A high quality polycrystalline sample (2 g) of SrCr$_2$As$_2$ was synthesized by solid-state reaction using Sr (99.95\%), Cr (99.99\%) and As (99.999 99\%) from Alfa-Aesar. The synthesis was started by reacting small pieces of Sr metal with prereacted CrAs taken in the ratio Sr:CrAs~=~1.05:2. Excess Sr was added in the starting composition to avoid the presence of unreacted CrAs phase and to compensate  for Sr loss due to evaporation. The mixture was pelletized, placed in an alumina crucible, and sealed in an evacuated quartz tube. The tube was placed in a box furnace and heated to 900~$^{\circ}$C  at a rate of 100~$^{\circ}$C/h and held at that temperature for 48 h, then the furnace was cooled to room temperature. This process was repeated twice with intermediate grinding. The resulting material was reground inside a helium-filled glove box, pelletized, and then sealed  under $\approx$ 1/4 atm high purity argon in a quartz tube. The sample was heated to 1150~$^{\circ}$C at the rate of 100~$^{\circ}$C/h and held there for 48 h followed by furnace cooling.  Powder x-ray diffraction of the final product confirmed the phase purity of  SrCr$_2$As$_2$. The magnetization $M$($T$) measurement in the temperature range 1.8 to 300~K was performed using a Quantum Design Inc., magnetic properties measurement system (MPMS). The high temperature $M$($T$) measurement from 300 to 900~K was performed using the vibrating sample magnetometer (VSM) option of a Quantum Design Inc., physical properties measurement system (PPMS).

Powder neutron diffraction measurements were performed at the thermal triple-axis spectrometer TRIAX at the University of Missouri Research Reactor. Measurements were carried out with an incident energy of 14.7 meV, using S\"{o}ller slit collimations of $60^\prime$-$60^\prime$-sample-$40^\prime$-$80^\prime$. Pyrolytic graphite filters were placed both before and after the sample to reduce higher-order wavelengths. The pelletized sample of mass $\sim$ 2 g was placed in an Al holder and was mounted on the cold finger (made of Cu) of a cryofurnace to reach temperatures of $12$~K~$\le T \le 612$~K. Rietveld refinements of the neutron diffraction data were carried out using FullProf software \cite{FullProf}.

\section{Results and Discussion}

\begin{figure}
\includegraphics[width=3.3in]{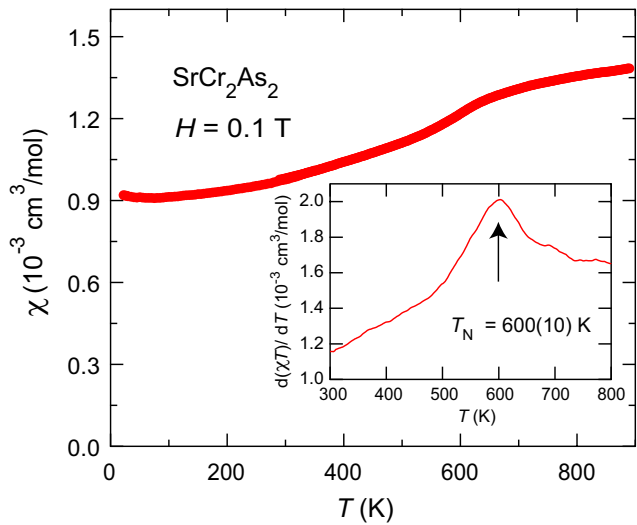}
\caption{(Color online) Zero-field-cooled (ZFC)  magnetic susceptibility, $\chi\equiv M/H$, of a  SrCr$_2$As$_2$ polycrystalline sample measured in the temperature ($T$) range of 1.8 to 900~K, at a magnetic field $H=0.1$~T. The inset shows d$(\chi T)$/d$T$ versus $T$, which yields the AFM ordering temperature \cite{Fisher1962} as $T_{\rm N}=~ $600(10)~K.}
\label{Fig:Magnetization}
\end{figure}

The temperature dependence of the magnetic susceptibility, $\chi\equiv M/H$, with an applied magnetic field $H=0.1$~T, is shown in Fig.~\ref{Fig:Magnetization}.  Over the extended temperature range, $\chi$ increases monotonically. The $\chi$($T$) shows a distinct  change in slope around $\sim 600$ K indicative of an AFM transition.  We identify the AFM transition temperature $T_{\rm N} = 600(10)$~K as the peak temperature of a $\lambda-$type anomaly obtained from d$(\chi T)$/d$T$ versus $T$ as shown in the inset of Fig.~\ref{Fig:Magnetization} \cite{Fisher1962}. We note that our $\chi$($T$) measurements and the neutron diffraction studies described below are inconsistent with the previous report of Ref. \cite{Pfisterer1983} suggesting an AFM transition at $T_{\textrm N}^{*}$ $\sim$~165~K which is evidently due to impurities. At temperatures above {\Tn}, the susceptibility appears to approach a broad maximum, indicative of strong two-dimensional AFM correlations setting in well above the ordering temperature, which by virtue of weak AFM interplanar coupling lead to the three dimensional AFM structure observed below {\Tn} \cite{Johnston2011,Vaknin1989}.

\begin{table*}[ht]
\caption{\label{Tab:FitParams} Fit parameters obtained from Rietveld refinements of the powder neutron diffraction patterns at the two listed temperatures with tetragonal $I4/mmm$ space group. $a$, $c$, and $V$ are the unit cell parameters and the unit cell volume, respectively. $z_{\rm As}$ represents the As $z$ position in the crystal structure. $d_{\rm Cr-Cr}$ and $d_{\rm Cr-As}$ are the in-plane Cr--Cr and Cr--As distances, respectively. $\chi^2$ gives the overall value of the goodness of fit. The error (one standard deviation) in the last digit of a quantity is shown in parentheses.}
\centering
\begin{ruledtabular}
\begin{tabular}{ccccccccc}
$T$ & $a$ & $c$ & $c/a$ & $V$ & $z_{\rm As}$ & $d_{\rm Cr-Cr}$ & $d_{\rm Cr-As}$ &  $\chi^2$ \\
(K) & ($\textrm{\AA}$) & ($\textrm{\AA}$) &	& ($\textrm{\AA}^3$) &  & ($\textrm{\AA}$) & ($\textrm{\AA}$) & \\
\hline
12  & 3.9063(8) & 12.933(4) & 3.311(1) & 197.35(8) & 0.3667(7) &  2.7622(6) & 2.468(3) & 3.04 \\
611 & 3.9619(7) & 12.921(4) & 3.261(1) & 202.82(8) & 0.3659(6) &  2.8015(5) & 2.483(2) & 2.90 \\

\end{tabular}
\end{ruledtabular}
\end{table*}

\begin{figure}
\includegraphics{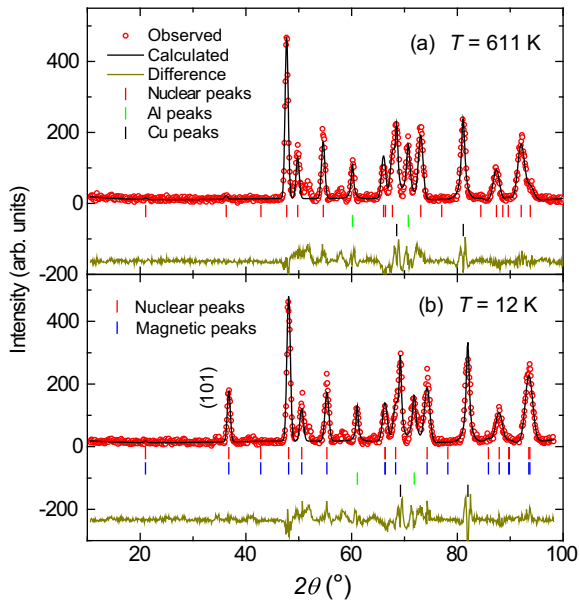}
\caption{(Color online)  Observed neutron diffraction patterns (red open circles), fits from the Rietveld refinement (black solid lines) and their differences (dark yellow solid lines) at (a) $T$~=~611~K ($>$~{\Tn}) and (b) $T$~=~12~K ($<$~{\Tn}). The vertical bars are the expected Bragg peak positions as mentioned in the panels. Additional peaks due to the Al sample holder and Cu cold finger were observed and have been taken into account while fitting.}
\label{Fig:DiffractionPattern}
\end{figure}

Figures~\ref{Fig:DiffractionPattern}(a) and \ref{Fig:DiffractionPattern}(b) show the full powder neutron diffraction pattern obtained at $T$~=~611~K ($>$~{\Tn}) and $T$~=~12~K ($<$~{\Tn}), respectively. Notice that all the nuclear and magnetic Bragg peaks coincide as shown in Fig.~\ref{Fig:DiffractionPattern}(b). No additional Bragg peaks are observed in the magnetically ordered state indicating the same chemical and magnetic unit cell, and furthermore that there is no structural phase transition down to 12~K. The magnetic intensities are superimposed on the nuclear Bragg peaks and decrease with increasing $2\theta$ in accordance with the expected behavior of a magnetic form factor. The strongest magnetic peak is the (1~0~1) Bragg reflection which is allowed by the chemical structure but has a very small nuclear structure factor.  Rietveld structural refinement of the nuclear structure at high temperature is performed using the tetragonal $I4/mmm$ \tcs-type crystal symmetry. The magnetic structure is determined from the combined nuclear and magnetic Rietveld refinements of the diffraction pattern at $T$~=~12~K, yielding a G-type AFM ordering with the magnetic Cr$^{2+}$ moments arranged antiferromagnetically with all nearest neighbors, both in-plane and out-of-plane, and aligned along the $c$ axis, as shown in Fig.~\ref{Fig:MagStructure}.  We note that the value of the ordered moment at $T$~=~12~K is found to be $\mu=1.9(1)~\mu_{\rm B}$/Cr, where $\mu_{\rm B}$ is the Bohr magneton, and is similar to \bca\ \cite{Filsinger2017}.

\begin{figure}
\includegraphics{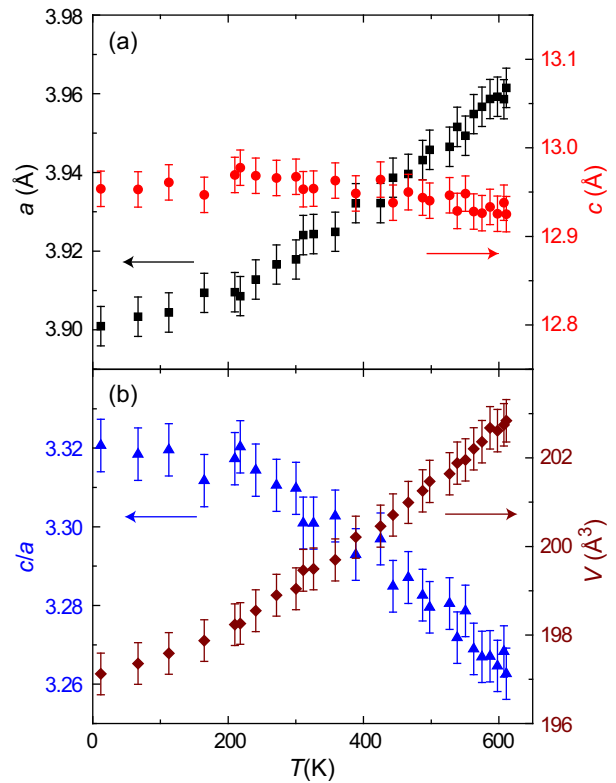}
\caption{(Color online)  (a) Lattice parameters $a$ and $c$ as a function of temperature, $T$. (b) $T$-dependence of the $c/a$ ratio and unit cell volume $V$.}
\label{Fig:LatticeParameter}
\end{figure}

The fit parameters from the Rietveld refinements of the diffraction patterns are listed in Table~\ref{Tab:FitParams}. The  lattice parameter $a$ and the unit cell volume $V=a^2c$ decrease by about $\sim$ 1.5\% and 2.5\%, respectively, between 611~K and 12~K, while the $c$ lattice parameter increases slightly. This is accompanied by an almost $\sim$ 1.5\% change in the Cr--Cr distance, compared with $\sim$ 0.5\% change in the Cr--As distance at the two temperatures.

\begin{figure}
\includegraphics{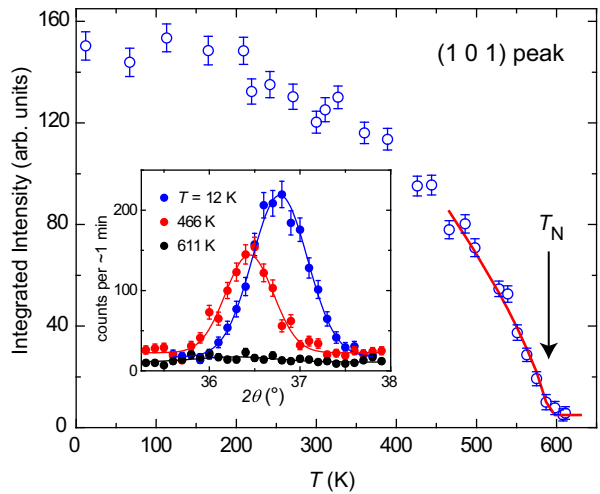}
\caption{(Color online) Integrated intensity ($I_{\textrm M}$) of the (1~0~1) Bragg peak as a function of temperature $T$. The solid line is a power law fit given by $I_{\textrm M}=I_0(1-T/T_{\textrm N})^{2\beta}$, for $T\ge 450$~K. Inset: $2\theta$ scans around the (1~0~1) Bragg peak for the temperatures listed.}
\label{Fig:OrderParameter}
\end{figure}

For temperature dependence measurements, two regions in $2\theta$ were chosen. The first region is centered around the (1~0~1) Bragg peak, $34^\circ \le 2\theta \le 39^\circ$, which has a weak nuclear contribution and for which the magnetic signal is the strongest, making it  ideal for the temperature dependence of the order parameter. The second region, $46^\circ \le 2\theta \le 52^\circ$, covers the (1~0~3) and (1~1~0) Bragg peaks, from which the temperature dependence of the lattice parameters and the unit cell volume were obtained. The  lattice parameter $a$ is obtained from the (1~1~0) Bragg peak and is then used to determine the  lattice parameter $c$ from the (1~0~3) Bragg peak. Figure~\ref{Fig:LatticeParameter}(a) shows the temperature dependence of the $a$ and $c$ lattice parameters while Fig.~\ref{Fig:LatticeParameter}(b) shows the temperature dependence of the $c/a$ ratio and the unit cell volume $V$. Since we do not have a full diffraction pattern at these temperatures, these are not Rietveld-refined values, but provide a good estimate obtained from the $2\theta$ values of the centers of the fitted Bragg peaks. The $a$ lattice parameter decreases monotonically from high temperatures while the $c$ lattice parameter remains almost constant throughout the measured temperature range with a slight increase with decreasing temperature. The $c/a$ ratio increases with decreasing temperature and becomes almost constant below 200~K while the unit cell volume $V$ decreases monotonically. These results are qualitatively similar to those in \bca\ \cite{Filsinger2017} but distinctly different from those of other Sr$T_2$As$_2$ ($T$ = Mn, Fe, Co) compounds. Specifically, \sma\  does not crystallize in a tetragonal $I4/mmm$ space group but forms a trigonal lattice with collinear AFM structure \cite{Das2017}, \sfa\ undergoes a first order structural transition from tetragonal to an orthorhombic AFM phase at low temperatures \cite{Li2009} and \scoa\ is non-magnetic with $I4/mmm$ crystal symmetry but undergoes a pressure-induced cT phase \cite{Jayasekara15}.

Figure~\ref{Fig:OrderParameter} shows the temperature dependence of the integrated intensity ($I_{\textrm M}$) of the (1~0~1) Bragg peak, which is a measure of the magnetic moment. The inset shows $2\theta$ scans of the (1~0~1) reflections at $T=$ 12~K, 466~K ($<$~{\Tn}) and 611~K ($>$~{\Tn}). As evident from the inset, the signal at 611~K is close to  background level as it has a negligible nuclear contribution. The shift in the peak position is due to the change in the lattice parameters with decreasing temperature. The continuous variation of the integrated intensity near {\Tn} indicates that the antiferromagnetic transition is thermodynamically of second order. For $T\ge 450$~K, we fitted the integrated intensity by a power law with a critical exponent $2\beta$, given by $I_{\textrm M}=I_0(1-T/T_{\textrm N})^{2\beta}$. From the fit, the antiferromagnetic transition temperature is found to be {\Tn} = 590(5)~K, which is, within error, consistent with  the transition temperature estimated from the $\chi$($T$) measurements in Fig.\ref{Fig:Magnetization}, {\Tn} = 600(10)~K. The critical exponent $\beta$ is found to be $\beta = 0.37(2)$, which is close to the expected value of 0.33 for a three-dimensional Heisenberg spin system. The intensity saturates below 200~K, which is also the same temperature below which the $c/a$ ratio becomes constant (see Fig.~\ref{Fig:LatticeParameter}), indicating a possible magneto-elastic coupling.

The small ordered moment, $\mu=1.9(1)~\mu_{\rm B}$/Cr, obtained from the Rietveld refinement, suggests that this is not a localized-moment system but rather an itinerant one. In particular, the full moment for a localized Cr$^{2+}$ is expected to be $\mu=gS\mu_{\rm B}=4\mu_{\rm B}$, assuming a ${^5}{\rm D}_0$ high-spin configuration in a tetrahedral environment with spin $S=2$ and spectroscopic splitting factor $g=2$, and is almost twice our experimental value. This itinerant character can be rationalized by strong hybridization between Cr $3d$ orbitals and the As $4p$ orbitals as has been suggested for \bca\ \cite{Singh2009a} and observed in the similar \bma\ compound \cite{Singh2009,Singh2009b}. From first-principle calculations, it is estimated that at the Fermi energy, the Cr $d$ orbitals contribute almost $2/3$ of the density of states while the remaining $1/3$ is of As $p$ character, resulting in large multi-sheet Fermi surfaces and making the system itinerant \cite{Singh2009a} with a significantly reduced ordered moment. Finally we come to the discussion of the magnetic exchange interactions $J_i$'s. In FeAs compounds like $A$Fe$_2$As$_2$ ($A$ = Ca, Ba, Sr), stripe-type AFM is stabilized with the Fe$^{2+}$ magnetic moments in the $ab$-plane.  It has been argued that  the stripe structure is driven by the  next-nearest neighbor (NNN) interaction term $J_2$ when  $J_2 \ge J_{1}/2$, where $J_1$ is the nearest neighbor (NN) interaction \cite{Yildirim2009,Han2009}. In our case of \sca, the G-type AFM suggests that NN interaction  $J_1$ is more dominant than $J_2$.

\section{Summary}

We have shown that \sca\ exhibits itinerant AFM with a G-type magnetic structure below {\Tn}~=~590(5)~K with the Cr magnetic moments aligned along the $c$ axis. However, strong magnetic correlations develop well above {\Tn} as evident from the susceptibility measurements. We find that the system remains tetragonal in the $I4/mmm$ symmetry down to the base temperature ($\sim$~12~K). The lattice parameter ratio $c/a$ and the ordered magnetic moment $\mu$ saturate at about the same temperature below $\sim$~200~K, indicating a possible magneto-elastic coupling. The derived $\mu=1.9(1)~\mu_{\rm B}$/Cr is significantly reduced due to the itinerant character of the system, caused by the hybridization between the Cr $3d$ and the As $4p$ orbitals.

\acknowledgments

This research was supported by the U.S. Department of Energy, Office of Basic Energy Sciences, Division of Materials Sciences and Engineering.  Ames Laboratory is operated for the U.S. Department of Energy by Iowa State University under Contract No.~DE-AC02-07CH11358.



\begin{thebibliography}{99}

\bibitem{Johnston2010} D. C. Johnston. The puzzle of high temperature superconductivity in layered iron pnictides and chalcogenides. Adv. Phys. {\bf 59}, 803 (2010).

\bibitem{Stewart2011} G. R. Stewart.  Superconductivity in iron compounds. Rev. Mod. Phys. {\bf 83}, 1589 (2011).

\bibitem{Scalapino2012} D. J. Scalapino.  A common thread: The pairing interaction for unconventional superconductors.  Rev. Mod. Phys. {\bf 84}, 1383 (2012).

\bibitem{Dagotto2013} E. Dagotto. The unexpected properties of alkali metal iron selenide superconductors.  Rev. Mod. Phys. {\bf 85}, 849 (1913).

\bibitem{Fernandes2014} R. M. Fernandes, A. V. Chubukov, and J. Schmalian. What drives nematic order in iron-based superconductors? Nat. Phys. {\bf 10}, 97 (2014).

\bibitem{Hosono2015} H. Hosono and K. Kuroki. Iron-based superconductors: Current status of materials and pairing mechanism. Physica C {\bf 514}, 399 (2015).

\bibitem{Dai2015} P. Dai. Antiferromagnetic order and spin dynamics in iron-based superconductors. Rev. Mod. Phys. {\bf 87}, 855 (2015).

\bibitem{Inosov2016} D. S. Inosov. Spin fluctuations in iron pnictides and chalcogenides: From antiferromagnetism to superconductivity. Compt. Rend. Phys. {\bf 17}, 60 (2016).

\bibitem{Si2016} Q. Si, R. Yu, and E. Abrahams. High-temperature superconductivity in iron pnictides and chalcogenides.  Nat. Rev. Mater. {\bf 1}, 1 (2016).

\bibitem{An2009} J. An, A. S. Sefat, D. J. Singh, and M.-H. Du. Electronic structure and magnetism in ${\rm BaMn_2As_2}$ and ${\rm BaMn_2Sb_2}$. Phys. Rev. B {\bf 79}, 075120 (2009).

\bibitem{Singh2009} Y. Singh, A. Ellern, and D. C. Johnston. Magnetic, transport, and thermal properties of single crystals of the layered arsenide ${\rm BaMn_2As_2}$.  Phys. Rev. B {\bf 79}, 094519 (2009).

\bibitem{Singh2009b}  Y. Singh, M. A. Green, Q. Huang, A. Kreyssig, R. J. McQueeney, D. C. Johnston, and A. I. Goldman. Magnetic order in BaMn$_2$As$_2$ from neutron diffraction measurements. Phys. Rev. B {\bf 80}, 100403(R) (2009).

\bibitem{Johnston2011}  D. C. Johnston, R. J. McQueeney, B. Lake, A. Honecker, M. E. Zhitomirsky, R. Nath, Y. Furukawa, V. P. Antropov, and Y. Singh. Magnetic exchange interactions in BaMn$_2$As$_2$: A case study of the $J_1$-$J_2$-$J_c$ Heisenberg model. Phys. Rev. B {\bf 84}, 094445 (2011).

\bibitem{Antel2012} A. Antal, T. Knoblauch, Y. Singh, P. Gegenwart, D. Wu, and M. Dressel. Optical properties of the iron-pnictide analog BaMn$_2$As$_2$.  Phys. Rev. B {\bf 86}, 014506 (2012).

\bibitem{Calder2014} S Calder, B. Saparov, H. B. Cao, J. L. Niedziela, M. D. Lumsden, A. S. Sefat, and A. D. Christianson. Magnetic structure and spin excitations in BaMn$_2$Bi$_2$.  Phys. Rev. B {\bf 89}, 064417 (2014).

\bibitem{Zhang2016} W.-L. Zhang, P. Richard, A. van Roekeghem, S.-M. Nie, N. Xu, P. Zhang, H. Miao, S.-F. Wu, J.-X. Yin, B. B. Fu, L.-Y. Kong, T. Qian, Z.-J. Wang, Z. Fang, A. S. Sefat, S. Biermann, and H. Ding. Angle-resolved photoemission observation of Mn-pnictide hybridization and negligible band structure renormalization in BaMn$_2$As$_2$ and BaMn$_2$Sb$_2$. Phys. Rev. B {\bf 94}, 155155 (2016).

\bibitem{Sangeetha2016} N. S. Sangeetha, A. Pandey, Z. A. Benson, and D. C. Johnston. Strong magnetic correlations to 900~K in single crystals of the trigonal antiferromagnetic insulators ${\rm SrMn_2As_2}$ and ${\rm CaMn_2As_2}$. Phys. Rev. B {\bf 94}, 094417 (2016).

\bibitem{Das2017} P. Das, N. S. Sangeetha, A. Pandey, Z. A. Benson, T. W. Heitmann, D. C. Johnston, A. I. Goldman, and A. Kreyssig. Collinear antiferromagnetism in trigonal ${\rm SrMn_2As_2}$ revealed by single-crystal neutron diffraction.  J. Phys.: Condens. Matter {\bf 29}, 035802 (2017).

\bibitem{Pfisterer1980} M. Pfisterer and G. Nagorsen. On the Structure of Ternary Arsenides. Z. Naturforsch. {\bf 35b}, 703 (1980).

\bibitem{Pfisterer1983} M. Pfisterer and G. Nagorsen. Bonding and Magnetic Properties in Ternary Arsenides $ET_2{\rm As}_2$.  Z. Naturforsch. {\bf 38b}, 811 (1983).

\bibitem{Singh2009a} D. J. Singh, A. S. Sefat, M. A. McGuire, B. C. Sales, D. Mandrus, L. H. VanBebber, and V. Keppens. Itinerant antiferromagnetism in ${\rm BaCr_2As_2}$: Experimental characterization and electronic structure calculations. Phys. Rev. B {\bf 79}, 094429 (2009).

\bibitem{Filsinger2017} K. A. Filsinger, W. Schnelle, P. Adler, G. H. Fecher, M. Reehuis, A. Hoser, J.-U. Hoffmann, P. Werner, M. Greenblatt, and C. Felser. Antiferromagnetic structure and electronic properties of ${\rm BaCr_2As_2}$ and ${\rm BaCrFeAs_2}$.  arXiv:1701.03127.

\bibitem{Nayak2017} J. Nayak, K. Filsinger, G. H. Fecher, S. Chadov, J. Minar, E. E. D. Rienks, B. B\"uchner, J. Fink, and C. Felser. Observation of a remarkable reduction of correlation effects in ${\rm BaCr_2As_2}$ by ARPES.  arXiv:1701.06108.

\bibitem{Richard2017} P. Richard, A. van Roekeghem, B. Q. Lv, T. Qian, T. K. Kim, M. Hoesch, J.-P. Hu, A. S. Sefat, S. Biermann, and H. Ding. Is ${\rm BaCr_2As_2}$ symmetrical to ${\rm BaFe_2As_2}$ with respect to half $3d$ shell filling? arXiv:1701.07591.

\bibitem{Naumov17} P. G. Naumov, K. Filsinger, O. I. Barkalov, G. H. Fecher, S. A. Medvedev, and C. Felser. Pressure-induced transition to the collapsed tetragonal phase in BaCr$_2$As$_2$. Phys. Rev. B {\bf 95}, 144106 (2017).

\bibitem{Quirinale13} D. G. Quirinale, V. K. Anand, M. G. Kim, Abhishek Pandey, A. Huq, P. W. Stephens, T. W. Heitmann, A. Kreyssig, R. J. McQueeney, D. C. Johnston, and A. I. Goldman. Crystal and magnetic structure of CaCo$_{1.86}$As$_2$ studied by x-ray and neutron diffraction. Phys. Rev. B {\bf 88}, 174420 (2013).

\bibitem{Goldman09} A. I. Goldman, A. Kreyssig, K. Proke\v{s}, D. K. Pratt, D. N. Argyriou, J. W. Lynn, S. Nandi, S. A. J. Kimber, Y. Chen, Y. B. Lee, G. Samolyuk, J. B. Le\~{a}o, S. J. Poulton, S. L. Bud'ko, N. Ni, P. C. Canfield, B. N. Harmon, and R. J. McQueeney. Lattice collapse and quenching of magnetism in CaFe$_2$As$_2$ under pressure: A single-crystal neutron and x-ray diffraction investigation. Phys. Rev. B {\bf 79}, 024513 (2009).

\bibitem{Jayasekara15} W. T. Jayasekara, U. S. Kaluarachchi, B. G. Ueland, Abhishek Pandey, Y. B. Lee, V. Taufour, A. Sapkota, K. Kothapalli, N. S. Sangeetha, G. Fabbris, L. S. I. Veiga, Yejun Feng, A. M. dos Santos, S. L. Bud'ko, B. N. Harmon, P. C. Canfield, D. C. Johnston, A. Kreyssig, and A. I. Goldman. Pressure-induced collapsed-tetragonal phase in SrCo$_2$As$_2$. Phys. Rev. B {\bf 92}, 224103 (2015).

\bibitem{Paramanik2014} U. B. Paramanik, R. Prasad, C. Geibel, and Z. Hossain. Itinerant and local-moment magnetism in ${\rm EuCr_2As_2}$ single crystals. Phys. Rev. B. {\bf 89}, 144423 (2014).

\bibitem{Nandi2016} S. Nandi, Y. Xiao, N. Qureshi, U. B. Paramanik, W. T. Jin, Y. Su, B. Ouladdiaf, Z. Hossain, and Th. Br\"uckel. Magnetic structures of the Eu and Cr moments in ${\rm EuCr_2As_2}$: Neutron diffraction study. Phys. Rev. B {\bf 94}, 094411 (2016).

\bibitem{Bao2015} J.-K. Bao, J.-Y. Liu, C.-W. Ma, Z.-H. Meng, Z.-T. Tang, Y.-L. Sun, H.-F. Zhai, H. Jiang, H. Bai, C.-M. Feng, Z.-A. Xu, and G.-H. Cao. Superconductivity in Quasi-One-Dimensional ${\rm K_2Cr_3As_3}$ with Significant Electron Correlations. Phys. Rev. X {\bf 5}, 011013 (2015).

\bibitem{Tang2015a} Z.-T. Tang, J.-K. Bao, Y. Liu, Y.-L. Sun, A. Ablimit, H.-F. Zhai, H. Jiang, C.-M. Feng, Z.-A. Xu, and G.-H. Cao. Unconventional superconductivity in quasi-one-dimensional ${\rm Rb_2Cr_3As_3}$. Phys. Rev. B {\bf 91}, 020506 (2015).

\bibitem{Tang2015b} Z.-T. Tang, J.-K. Bao, Z. Wang, H. Bai, H. Jiang, Y. Liu, H.-F. Zhai, C.-M. Feng, Z.-A. Xu, and G.-H. Cao. Superconductivity in quasi-one-dimensional ${\rm Cs_2Cr_3As_3}$ with large interchain distance. Sci. Chin. Mater. {\bf 58}, 16 (2015).

\bibitem{Wu2014} W. Wu, J. Cheng, K. Matsubayashi, P. Kong, F. Lin, C. Jin, N. Wang, Y. Uwatoko, and J. Luo. Superconductivity in the vicinity of antiferromagnetic order in CrAs.
    Nat. Commun. {\bf 5}, 5508 (2014).

\bibitem{Kotegawa2014} H. Kotegawa, S. Nakahara, H. Tou, and H. Sugawara. Superconductivity of 2.2K under Pressure in Helimagnet CrAs. J. Phys. Soc. Jpn. {\bf 83}, 093702 (2014).

\bibitem{Sangeetha2017} N. S. Sangeetha and D. C. Johnston (unpublished).

\bibitem{FullProf} J. Rodriguez-Carvajal. Recent advances in magnetic structure determination by neutron powder diffraction. Physica B {\bf 192}, 55 (1993).

\bibitem{Li2009} Haifeng Li, Wei Tian, Jerel L. Zarestky, Andreas Kreyssig, Ni Ni, Sergey L. Bu'ko, Paul C. Canfield, Alan I. Goldman, Robert J. McQueeney, and David Vaknin. Magnetic and lattice coupling in single-crystal SrFe$_2$As$_2$: A neutron scattering study. Phys. Rev. B {\bf 80}, 054407 (2009).

\bibitem{Fisher1962} M. E. Fisher. Relation between the specific heat and susceptibility of an antiferromagnet. Phil. Mag. {\bf 7}, 1731 (1962).

\bibitem{Vaknin1989} D. Vaknin, E. Caignol, P. K. Davies, J. E. Fischer, D. C. Johnston, and D. P. Goshorn. Antiferromagnetism in (Ca$_{0.85}$Sr$_{0.15}$)CuO$_2$, the parent of the cuprate family of superconducting compounds. Phys. Rev. B {\bf 39}, 9122 (1989).

\bibitem{Yildirim2009} T. Yildirim. Frustrated magnetic interactions, giant magneto–elastic coupling, and magnetic phonons in iron–pnictides. Physica C {\bf 469}, 425 (2009).

\bibitem{Han2009} M. J. Han, Q. Yin, W. E. Pickett, and S. Y. Savrasov. Anisotropy, Itineracy, and Magnetic Frustration in High-$T_{\rm C}$ Iron Pnictides. Phys. Rev. Lett. {\bf 102}, 107003 (2009).






\end{thebibliography}
\end{document}